\documentclass{camera_1}
\usepackage{graphicx}  
\usepackage[latin1]{inputenc}

\begin{document}

%
\title{Alpha-particle clustering in excited expanding self-conjugate
nuclei}

%
\author{B.~Borderie$^1$, Ad.~R.~Raduta$^{1,2}$, G.~Ademard$^1$,
M.~F.~Rivet$^1$, E.~De~Filippo$^3$, E.~Geraci$^3$,
N.~Le~Neindre$^{1,4}$, G.~Cardella$^3$, G.~Lanzalone$^5$,
I.~Lombardo$^5$, O.~Lopez$^{4}$, C.~Maiolino$^5$, A.~Pagano$^3$,
S.~Pirrone$^3$, G.~Politi$^3$, F.~Rizzo$^{5,3}$ 
\and P.~Russotto$^{5,3}$}

%
\organization{$^1$ Institut de Physique Nucl\'eaire, CNRS/IN2P3, Universit\'e Paris-Sud 11,
91406 Orsay, France \\
$^2$ National Institute for Physics and Nuclear Engineering,
Bucharest-Magurele, Romania\\
$^3$ INFN, Sezione di Catania and Dipartimento di Fisica e
Astronomia, Universit\`a di Catania, Italy\\
$^4$ LPC, CNRS/IN2P3, Ensicaen, Universit\'e de Caen, Caen, France\\
$^5$ INFN, Laboratori Nazionali del Sud, Catania, Italy}
\maketitle

\begin{abstract}
The fragmentation of quasi-projectiles from the nuclear reaction
$^{40}$Ca+$^{12}$C at 25 MeV/nucleon was used to produce
$\alpha$-emission sources. From a careful selection of these sources
provided by a complete detection and
from comparisons with models of sequential and simultaneous decays,
strong indications in favour of $\alpha$-particle clustering in excited $^{16}O$,
$^{20}Ne$ and $^{24}Mg$ are reported. 
\end{abstract}

\section{Introduction}
Clustering is a generic phenomenon which can appear in homogeneous matter when density
decreases; the formation of galaxies as well as the disintegration
of hot dilute heavy nuclei into lighter nuclei are extreme examples occuring in nature.
As far as nuclear physics is concerned, the nucleus viewed as a collection of
$\alpha$-particles was very early
discussed and in the last forty years both
theoretical and experimental efforts were devoted to clustering  phenomena in
nuclei. Very recently the formation of $\alpha$-particle clustering from
excited expanding self-conjugate
nuclei was revealed in two different constrained self consistent mean
field calculations~\cite{girod_prl2013,ebran_2014}.
The aim of the present work was to obtain, from the experimental side,
some information on $\alpha$-particle clustering from excited
and consequently expanding alpha-conjugate nuclei.
The chosen experimental strategy was to use
the reaction $^{40}$Ca+$^{12}$C at an incident energy (25 MeV per
nucleon) sufficient to possibly produce some hot expanding reaction
products, associated with a
high granularity-high solid angle particle array (to precisely reconstruct
directions of velocity vectors). Then, by selecting the appropriate
reaction mechanism and specific events 
the required information was derived.
   
\section{Experimental details and event selection}
The experiment was performed at INFN,
Laboratori Nazionali del Sud in Catania, Italy. 
The beam impinging on a thin carbon target (320 $\mu$g/cm$^2$)
was delivered by the Superconducting Cyclotron and the
charged reaction products were detected by the CHIMERA 4$\pi$
multi-detector~\cite{chimera}. The beam intensity was kept around
$10^7$ ions/s to avoid pile-up events. 
CHIMERA consists of 1192 telescopes ($\Delta$E silicon detectors
200-300 $\mu$m thick and CsI(Tl) stopping
detectors) mounted on 35 rings covering 94\% of the solid angle,
with very high granularity at forward angles.
Details on A and Z identifications and on the quality of
energy calibrations can be found in
refs.~\cite{chimera,leneindre_nim2002,adriana_plb}. Energy
resolution was better than 1\% for silicon detectors and varies
between 1.0 and 2.5\% for alpha particles stopped in CsI(Tl) crystals. 

\begin{figure}
    \begin{minipage}[t]{0.49\textwidth}
	\includegraphics[width=\textwidth]{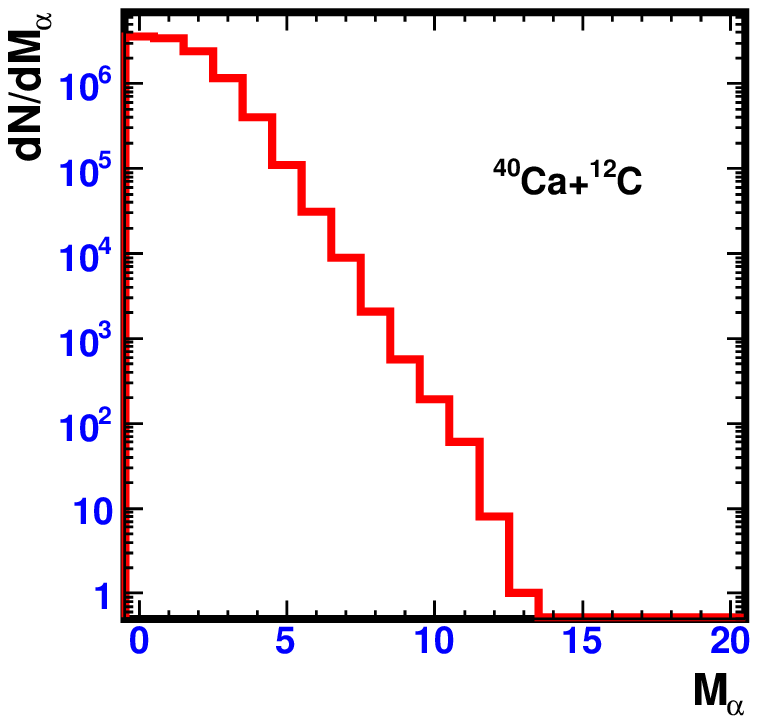}
	\caption{Distribution of $\alpha$-particle multiplicity, $M_{\alpha}$, for well detected
         events ($Z_{tot}\geq$19) } \label{fig:Malpha}
    \end{minipage}%
    \hspace*{0.02\textwidth}%
    \begin{minipage}[t]{0.49\textwidth}
	\includegraphics[width=\textwidth]{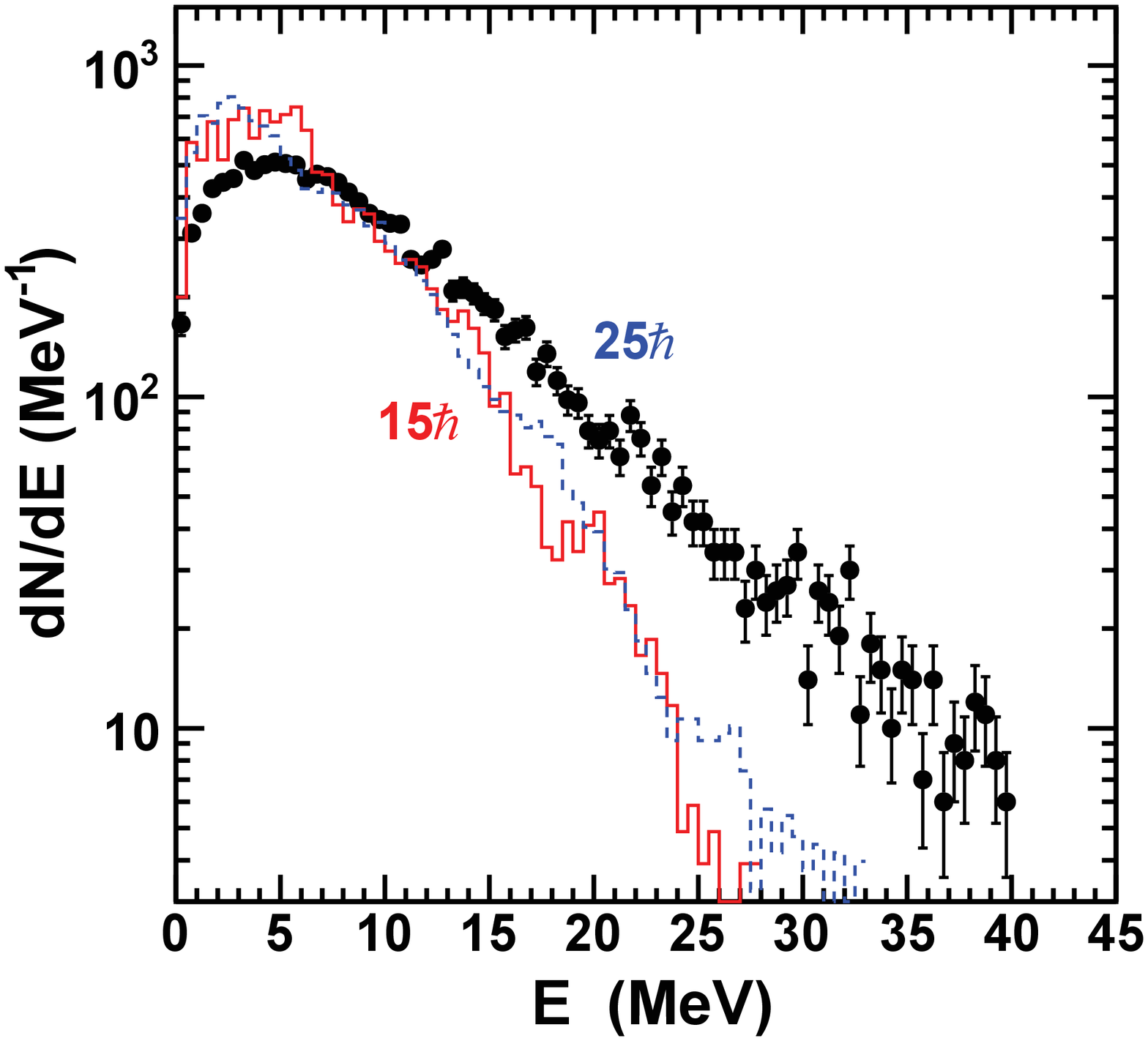}
	\caption{Sequential decay of excited $Ca$ projectiles: 
	energy spectra (in the $N\alpha$=5 system reference frame) of
	evaporated $\alpha$-particles associated to a  $^{20}Ne$
	evaporation residue . Full points are experimental data and
	histograms are results of GEMINI simulations (see text)}
	\label{fig:SEQPF}
    \end{minipage}
\end{figure}

As a first step in our event selection procedure, we want to exclude
from the data sample poorly-measured events. Without making any
hypothesis about the physics of the studied reaction one can measure
the total detected charge $Z_{tot}$ (neutrons are not measured).
In relation with their cross-sections  and with the
geometrical efficiency of CHIMERA, the well detected reaction mechanisms 
correspond to projectile
fragmentation (PF)~\cite{morjean,profrag,LTG} with $Z_{tot}$=19-20
(target fragmentation not
detected) and to incomplete/complete fusion with $Z_{tot}$=21-26.
At this stage we can have a first indication on the multiplicity
of $\alpha$-particles, $M_{\alpha}$, emitted per event for 
well detected events ($Z_{tot}\geq$19 - see fig.\ref{fig:Malpha}).
$M_{\alpha}$ extends up to thirteen, which means a
deexcitation of the total system into $\alpha$-particles only. Moreover a
reasonable number of events exhibit $M_{\alpha}$ values up to about 6.

The goal is now to tentatively isolate, in events, reaction products emitting
$\alpha$-particles only. To do this, knowing that at such incident
energy $^{40}Ca$ as $^{20}Ne$ PF is dominated by alpha-conjugate
products~\cite{morjean}, we
restrict our selection to completely detected PF events ($Z_{tot}$=20)
composed of one projectile fragment and of four to six $\alpha$-particles.
Charge conservation imposes $Z_{frag}$=20 - 2$M_{\alpha}$.

After this double selection, the question is: from which emission
source are the $\alpha$-particles emitted?
Several possibilities have to be considered and complementary
selections must be done before
restricting our study to alpha-sources emitting exclusively the $M_{\alpha}$
observed (called $N\alpha$ sources in what follows).
Possibilities associated to selected PF events are the following:

 I) considering the incident energy of the reaction and the strong forward
  focusing of reaction products, it is important to identify
  the possible presence of preequilibrium (PE) $\alpha$-particles in
  our selected PF events. With the hypothesis that all the 
  $\alpha$-particles are emitted from their centre-of-mass reference
  frame, we noted a thermal distribution
  with the presence of a high energy
  tail starting at 40 MeV, which signs PE emission. 
  Events in which such PE emission was possibly present were suppressed
  to prevent errors on alpha emitter properties;
  an upper energy limit was imposed to the $\alpha$-particle energy,
  with a value of 40 MeV irrespective of $M_{\alpha}$.
  
 II) $\alpha$-particles can be emitted from deexcitation of  PF events via
  unbound states of $^{12}C$, $^{16}O$, $^{20}Ne$  and not directly from
  excited expanding $N\alpha$ sources. Multi-particle correlation
  functions~\cite{charity_prc1995,adriana_plb} were used to identify
  unbound states 100\% $\alpha$-particle
  emitters and suppress a small percentage of events (1.6-3.6\%).

III) it must be verified that the fragments associated with
  $M_{\alpha}$ are not the evaporation residues of excited $Ca$
  projectiles emitting only $\alpha$-particles.  

As far as the two first items are concerned the effect was to
suppress from 8.5 ($M_{\alpha}$=4) to 12.8\% ($M_{\alpha}$=6)
of previously selected events. The last item will be discussed in the
following section.

To conclude on this part, one can also indicate that if excited $N\alpha$
sources have been formed their excitation energy
thresholds for total deexcitation into $\alpha$-particles vary from 20
to 50 MeV when $N\alpha$ moves from 4 to 6. Their mean excitation
energy per nucleon is rather constant around 3.3-3.5 MeV which
indicates that average lowest densities around 0.7 the normal density
may have been reached due to thermal
pressure~\cite{friedman_prc,borderie_brolo}.

\section{Evidence for $\alpha$-particle clustering}
Before discussing different possible deexcitations of $Ca$ projectiles
and $N\alpha$ sources,
information on the selected reaction mechanism is needed. 
Major features of PF events 
are reproduced by a model of stochastic transfers~\cite{LTG}. For
primary events with $Z_{tot}$=20 excitation energy
 extends up to about 200 MeV, which 
allows the large excitation energy domain
(20-150 MeV) measured for $N\alpha$ sources when associated to a single
fragment; angular momenta extend up to 24 $\hbar$, which gives
an upper spin limit for $Ca$ projectiles or $N\alpha$ sources.

Are $\alpha$-particles emitted sequentially or simultaneously?
To answer the question $\alpha$-energy spectra are compared to
simulations. For excited $Ca$ projectiles and $N\alpha$ sources, 
experimental velocity and excitation
energy distributions and distributions for spins are used as
inputs. Then, results of simulations are
filtered by the multi-detector replica.
Simulated spectra are
normalized to the area of experimental spectra. 

For sequential emission the GEMINI++ code~\cite{gemini} was used.
Before discussing $N\alpha$ sources, as said before, we must
consider the possible evaporation from $Ca$ projectiles.
Excitation energy for projectiles is deduced from 
$E^*$=$E^{*}(N\alpha)$+$E_{rel}$+$Q$.
$E_{rel}$ is the relative energy between the $N\alpha$ source and the
associated fragment (evaporation residue). Fig. \ref{fig:SEQPF} displays results of simulations
with reconstructed excitation energy distribution for
$^{40}Ca$ ($<E^*>$=88.2 MeV) and gaussian distributions centred at 15
and 25$\hbar$ for spins
(RMS=1.5$\hbar$) as inputs; note that no more $^{20}Ne$ residues are
produced for spin distributions centred at values larger than 25$\hbar$.
Comparison with experimental data shows a rather poor agreement indicating that
such an hypothesis seems not correct. Same kind of results are observed for
$N\alpha$ equal 4 and 6. 
   \begin{figure}
       \begin{minipage}[t]{0.49\textwidth}
	  \includegraphics[width=\textwidth]{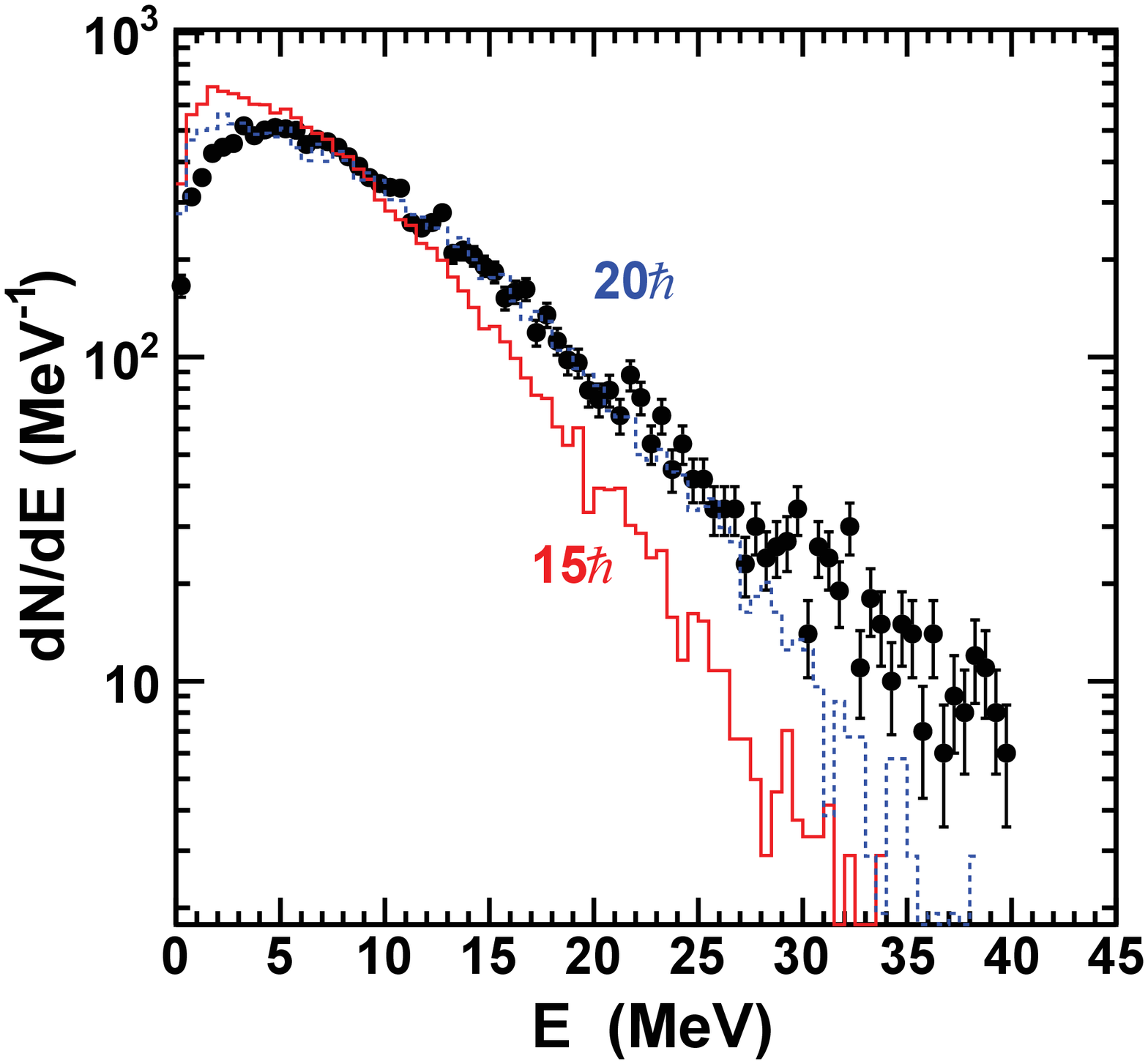}
	  \caption{Decay of excited $^{20}Ne$ ($N\alpha$=5): $\alpha$-particle energy spectra; 
	  full points are experimental data and histograms correspond to GEMINI
          simulations (see text).} \label{fig:EA5seq}
	\end{minipage}%
	\hspace*{0.02\textwidth}%
	\begin{minipage}[t]{0.49\textwidth}
	  \includegraphics[width=\textwidth]{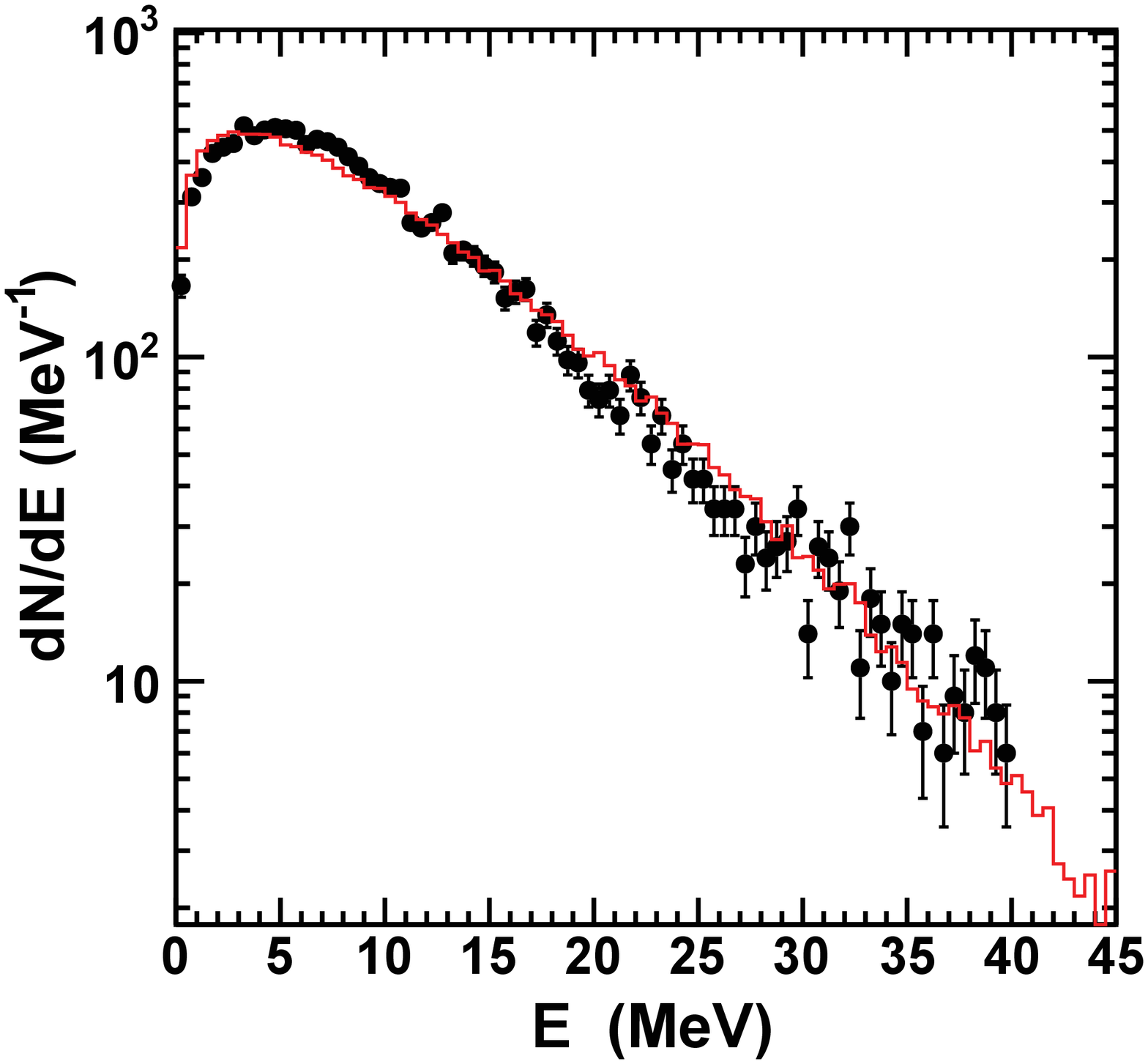}
	  \caption{same as fig. \ref{fig:EA5seq} but histogram
	  corresponds to a simulation of a simultaneous decay (see text).} \label{fig:EA5simul}
	\end{minipage}
   \end{figure}

Considering now excited $N\alpha$ sources only,
histograms in fig.~\ref{fig:EA5seq} are examples of GEMINI simulation
results for $N\alpha$=5.
Gaussian distributions for spins are used as
inputs and the best agreement with data is obtained with RMS=1.5$\hbar$
for spin distributions.
The agreement between 
data and simulations appears poorer and poorer when $N\alpha$ value decreases.
Moreover an important disagreement
between data and simulations
is observed for the percentages of $N\alpha$ sources which
deexcite via $^8Be$ emission~\cite{borderie_plb}.

For simultaneous emission from $N\alpha$ sources, a simulation was done which
mimics a situation in which $\alpha$ clusters are early formed when the
source is expanding~\cite{girod_prl2013,ebran_2014} due to thermal
pressure. The $N\alpha$ source is first splitted into $\alpha$'s.
Then the remaining available energy ($E^* + Q$) is directly randomly
shared among the $\alpha$-particles such as to conserve energy and linear
momentum.
Histogram in fig. \ref{fig:EA5simul} is the result of such a simulation,
which shows a good agreement with data. Such agreement is also observed
for $N\alpha$ equal 4 and 6.
Similar histograms (within a few
percents) were also obtained with simulations containing an intermediate freeze-out
volume stage and then propagation in the Coulomb field.

\section{Conclusions}
In conclusion, the reaction $^{40}$Ca+$^{12}$C at 
25 MeV/nucleon bombarding energy was used to produce and
carefully select minor classes of events from which excited
$N\alpha$ sources can be unambiguously identified.
Their excitation energy distributions are derived with mean values
around 3.4 MeV per
nucleon, which indicates that mean densities about 0.7 the normal density
may have been reached.
Their energetic emission properties were compared with two simulations,
one involving sequential decays and a second for simultaneous decay.
For excited expanding $N\alpha$ sources composed of 4, 5 and 6 $\alpha$-particles,
for which statistics is good enough for conclusives comparisons with
simulations,
evidence in favour of simultaneous emission
($\alpha$-particle clustering) is reported.

%
%

\end{document}